\documentclass[conference]{IEEEtran}
\IEEEoverridecommandlockouts

\usepackage{cite}
\usepackage[colorlinks]{hyperref}
\hypersetup{citecolor=blue}
\usepackage{amsmath,amssymb,amsfonts}
\usepackage{algorithmic}
\usepackage{algorithm}

\usepackage{graphicx}
\usepackage[font=footnotesize]{caption}
\usepackage[font=footnotesize]{subcaption}
\usepackage{placeins}
\usepackage{textcomp}
\usepackage{float}
\usepackage{booktabs}
\usepackage{xcolor}
\usepackage{balance}
\usepackage{fancyhdr}

\def\BibTeX{{\rm B\kern-.05em{\sc i\kern-.025em b}\kern-.08em
		T\kern-.1667em\lower.7ex\hbox{E}\kern-.125emX}}

	\usepackage[font=small,skip=2pt]{caption}

	\usepackage{enumitem}
	\setlist{nosep}
\begin{document}
	\title{\LARGE Fluid Antenna-assisted Unsourced ISAC Massive Access}
	\author{Jingyuan Xu, Zhentian Zhang, Hao Jiang, Jian Dang, Zaichen Zhang

		\thanks{ }
	\thanks{Jingyuan Xu, Zaichen Zhang and Zhentian Zhang are with the National Mobile Communications Research Laboratory, Frontiers Science Center for Mobile Information Communication and Security, Southeast University, Nanjing, 210096, China. Zaichen Zhang is also with the Purple Mountain Laboratories, Nanjing 211111, China. (e-mails: \{213223473, zczhang\}@seu.edu.cn, zhentianzhangzzt@gmail.com).

		Jian Dang is with the National Mobile Communications Research Laboratory, Frontiers Science Center for Mobile Information Communication and Security, Southeast University, Nanjing 211189, China, also with the Key Laboratory of Intelligent Support Technology for Complex Environments, Ministry of Education, Nanjing University of Information Science and Technology, Nanjing 210044, China, and also with Purple Mountain Laboratories, Nanjing 211111, China. (email: dangjian@seu.edu.cn).

		H. Jiang is with the School of Cyber Science and Engineering, Southeast University, Nanjing 210096, China. (e-mail: jiang.hao@seu.edu.cn).}
	\thanks{This work of Jian Dang and Zaichen Zhang is partly supported by the Fundamental Research Funds for the Central Universities (2242022k60001), Basic Research Program of Jiangsu (No. BK20252003), the Key Laboratory of Intelligent Support Technology for Complex Environments, Ministry of Education, Nanjing University of Information Science and Technology (No. B2202402). The work of Hao Jiang is supported in part by the National Natural Science Foundation of China (NSFC) projects (No. 62471238).}
	}
\maketitle
\pagestyle{fancy}
\thispagestyle{fancy}
\begin{abstract}
Unsourced integrated sensing and communication (UNISAC) has emerged as a promising paradigm for supporting massive connectivity in 6G networks. However, existing approaches predominantly rely on fixed-position antennas at the base station (BS) and user equipment (UE). In uplink transmission with huge access density and limited resource budgets (i.e., finite blocklength, FBL), the fixed arrays are constrained by their physical aperture and static spatial sampling, which lead to severe multi-user interference and an unavoidable pilot collision error floor. To conquer the bottleneck derived from fixed-position physical constraint and utilize the abundant spatial diversity within compact space, this paper proposes a novel unsourced ISAC framework incorporating a fluid antenna system (FAS) at the user side. The proposed scheme exploits the positional flexibility of FAS to reconfigure the channel environment by continuously adjusting antenna ports in the spatial domain. Numerical results demonstrate that the proposed FAS-aided approach significantly reduces the per-user probability of error (PUPE) and enhances angle-of-arrival (AOA) sensing accuracy. Specifically, the proposed scheme provides a 40 dB capacity gain over traditional TDMA at 1000 active users. It should be noted that the FAS considered in this paper is only deployed at the transmitter. In our future work, we will try deploying FAS at both the transmitter and receiver.
\end{abstract}
\begin{IEEEkeywords}
Unsourced integrated sensing and communication, finite blocklength, fluid antenna system
\end{IEEEkeywords}
\section{Introduction}
\subsection{Background: Massive Connectivity}
In the evolutionary progress to sixth-generation (6G) wireless communication networks, the underlying logic of global communication infrastructure is undergoing a paradigm shift from individual broadband communications to Internet-of-Everything-enabled massive machine-type communications (mMTC) \cite{mMTC}, which is rooted in information-theoretic perspectives, such as finite blocklength (FBL) transmission \cite{FBL}. With the comprehensive deployment of industrial Internet of Things (IoT) \cite{IOT1,IOT2}, autonomous driving networks, real-time digital twins, and smart city grids, future 6G networks must possess the capability to provide ultra-high reliability, ultra-low latency, and extremely high spectral efficiency in ultra-dense device deployment environments. Existing multiple access architectures are confronted with fundamental theoretical and physical bottlenecks.
\subsection{Related Work: Unsourced ISAC and FAS}
To fundamentally address the signaling overhead catastrophe and finite blocklength constraints associated with massive node access, unsourced random access (URA) has been proposed and rapidly emerged as a core enabling technology for 6G massive access \cite{URA0,URA1,URA1.2,URA1.3,URA1.4}, which completely abandons the physical-layer binding between user identity and transmitted data. In the URA architecture, all users in the network share a single global common codebook. When a device is activated and needs to transmit data, it directly maps its information bits to and selects a codeword from this common codebook for transmission. For the base station receiver, the decoding task transforms from the traditional ``\textit{identifying which specific user sent what information}" into a pure coding problem, ``\textit{which codewords in the massive common codebook have been activated and transmitted}." This turns the multiple access problem into an approximate support recovery and compressed sensing problem \cite{URA2}. Moreover, for quasi-static fading channels, efficient transmission strategies have been proposed to enhance system robustness. Theoretical derivations and simulations demonstrate that by deploying large-scale antenna arrays at the base station, URA systems exhibit remarkable energy efficiency and multi-user capacity.

Another core technological vision for 6G networks is known as integrated sensing and communication (ISAC) \cite{ISAC0,UNISAC2,ISAC2,ISAC22,tradeoffs}. ISAC systems achieve mutual assistance and gains between communication and sensing through shared spectrum, hardware radio-frequency chains, and unified waveform design. To reduce overhead, researchers have proposed the unsourced ISAC (UNISAC) models \cite{UNISAC1,UNISAC2,UNISAC3,UNISAC4,UNISAC5}. This model inherits the elegant coding-theoretic properties of URA, enabling simultaneous blind decoding of communication data for thousands of active users and high-precision estimation of sensing parameters, such as angle-of-arrival (AOA), under extremely low signal-to-noise ratios \cite{BXU1,BXU2} and finite blocklength conditions.

To thoroughly break the physical bottlenecks imposed by fixed antenna arrays, fluid antenna systems (FAS) \cite{Tutorial,FAS4,FAS1,FAS2,crb_AD,FAA} have been invented, which can provide lower error rate and better angle estimation \cite{FAS3}. This is particularly the case in massive access scenarios, an area known as fluid antenna multiple access (FAMA) \cite{FAMA1,FAMA2}. In the related research, a typical single-antenna FAS is modeled as having $N$ preset ``ports" densely distributed within a one-dimensional linear space of length $W \lambda$  (where $\lambda$ denotes the wavelength). Since the spacing between these ports is significantly smaller than the conventional half-wavelength constraint, the system can instantaneously switch the RF chain to one or several ports with optimal channel fading conditions through software control, achieving performance far exceeding traditional MIMO systems.
\subsection{Contributions}
To address the physical aperture constraints and static spatial sampling of conventional fixed arrays in massive access scenarios, we propose a novel unsourced ISAC framework incorporating FAS at the user side. The proposed framework adopts a slotted two-phase transmission architecture, where each time chunk consists of a compressed sensing (CS) phase for pilot transmission, activity detection, channel estimation, AOA estimation and forward error correction (FEC) phase decoding. At the user side, each UE is equipped with a linear fluid antenna that continuously adjusts its activated port position to select the channel with the maximum gain. At the BS side, a joint decoder integrating simultaneous orthogonal matching pursuit (SOMP), estimation of signal parameters via rotational invariance techniques (ESPRIT)-based AOA estimation, channel estimation refinement, and an alternating successive interference cancellation (SIC) strategy is designed to recover communication messages and sensing parameters. Numerical results demonstrate that the proposed FAS-assisted scheme significantly reduces the per-user probability of error (PUPE) and enhances AOA estimation accuracy. Specifically, the proposed scheme delivers a 40~dB capacity gain over conventional TDMA with 1000 active users.

The remainder of the paper is organized as follows: Sec. \ref{sec.2} presents the system model of unsourced ISAC and FAS. Sec. \ref{sec.3} discusses the design of the unsourced ISAC channel coding. In Sec. \ref{sec.4}, we present numerical results, where the proposed FAS scheme is evaluated under various parameter configurations to demonstrate its viability. Finally, Sec. \ref{sec.5} concludes the paper.
\section{System Model}\label{sec.2}
\begin{figure}[htbp]
	\centerline{\includegraphics[width=0.4\textwidth]{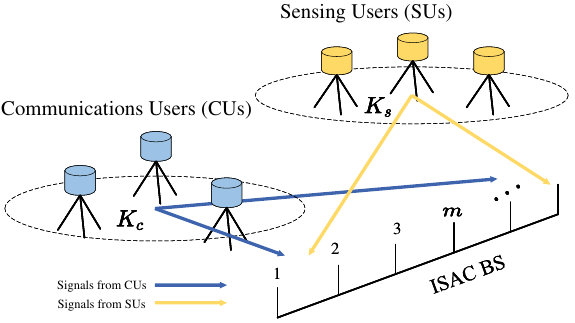}}
	\caption{Illustration of the unsourced ISAC system.}
	\label{system_model}
\end{figure}

This work considers the uplink transmission of unsourced ISAC, where a base station (BS) receives signals from $K_a$ active single-antenna user equipments (UEs) including $K_c$ communication users (CUs) and $K_s$ sensing users (SUs), i.e., $K_a= K_c+K_s$. For CUs, they intend to send $B_c$ bits of information and we use $\mathcal{L}_c$ to denote the message list of CUs, i.e., $|\mathcal{L}_c|=K_c$. For SUs, they generate signals via $B_s$ bits and we use set $\mathcal{L}_s=\{\theta_k\}, k\in[1:K_s]$ to denote the AOA. Omitting errors from asynchronous transmission, all UEs share identical finite transmission duration of $L$ channel uses. We illustrate the basic structure of unsourced ISAC in Fig.~\ref{system_model}.
\begin{figure}[htbp]
	\centerline{\includegraphics[width=0.35\textwidth]{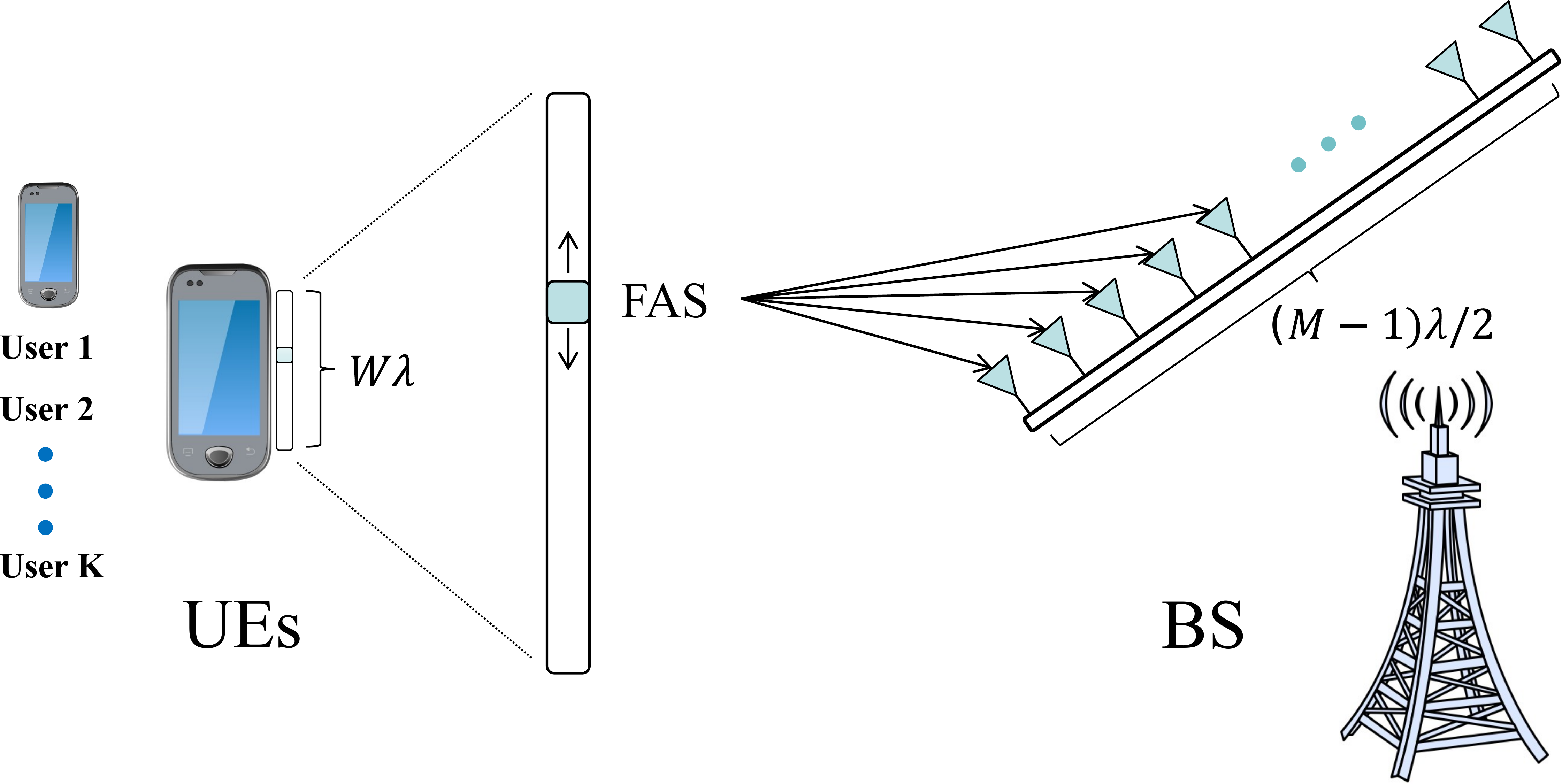}}
	\caption{Illustration of the fluid antenna system.}
	\label{FAS_in_UE}
\end{figure}
\subsection{Unsourced ISAC System Descriptions}
For unsourced ISAC, the BS aims to restore the binary messages of CUs and estimate the AOA of SUs. Aligning the paradigm in \cite{ISAC2}, the system performance is evaluated in terms of PUPE $\epsilon$ and the mean-square error of AOA (MSEAOA), defined as:
\begin{equation}
	\label{eq:1}
	\begin{aligned}
		\mathrm{PUPE} &=\frac{\mathbb{E}\{\mathcal{L}_{c,e}+\mathcal{L}_{s,e}\}}{K_c+K_s}, \\
		\mathrm{MSEAOA} &=\frac{1}{|\mathcal{L}_{s,d}|}\sum_{\theta_k\in \mathcal{L}_{s,d}}\mathbb{E}\{|\cos\theta_k-\cos\hat{\theta}_k|^2\},
	\end{aligned}
\end{equation}
where sets $\mathcal{L}_{c,e}$ and $\mathcal{L}_{s,e}$ denote the detection error of communication and sensing tasks, set $\mathcal{L}_{s,d}$ denotes the AOA of successfully detected SUs and $\hat{\theta}_k$ is the corresponding AOA estimation. The power of unsourced ISAC is indicated by the energy-per-user defined as:
\begin{equation}
	\label{eq:2}
	\frac{E}{N_0}=\frac{P}{K_a\sigma^2},
\end{equation}
where the numerator $P$ denotes the total power and $\sigma^2$ in the denominator is the noise variance of the additive white Gaussian noise (AWGN). Let $P_c$ and $P_s$ denote the power allocated to single CU and SU and we use $\alpha$ to denote the ratio of power for communication purpose, i.e., $P_c=\alpha P/K_c$ and $P_s=(1-\alpha) P/K_s$.
\subsection{FAS Channel Model}
The base station is equipped with $M$ fixed antennas, with equal spacing of $\lambda/2$. Let the incident angle be $\theta_k$, and the $m$-th channel vector can be expressed as:
\begin{equation}
	\boldsymbol{h}(m) = e^{-j\pi (m-1)\cos\theta_k}, \quad m \in [1:M].
\end{equation}

Meanwhile, each user is equipped with a linear FAS of size $W \lambda$, and the activation position can be selected continuously, whose structure is shown in Fig.~\ref{FAS_in_UE}. The channel from each port to the base station follows a complex Gaussian distribution, with channel vector $\boldsymbol{g}_0$ following distribution of $\mathcal{CN}(0, \sigma_h^2 \boldsymbol{I})$. For a linear FAS, assume that the position of $m$-th port is $d_m$, the position of $n$-th port is $d_n$. The correlation coefficient of the two ports can be expressed as:
\begin{equation}
\boldsymbol{\Sigma}_{m,n} = J_0\left(2\pi \frac{|d_m-d_n|}{\lambda}\right).
\end{equation}
We decompose this correlation matrix as:
\begin{equation}
\boldsymbol{\Sigma} = \boldsymbol{Q}\boldsymbol{\Lambda} \boldsymbol{Q}^{\mathrm H}.
\end{equation}
For each user, the channel gain vector of $N_a$ ports is equal to
\begin{equation}
	\boldsymbol{g} = \boldsymbol{Q} \sqrt{\boldsymbol{\Lambda}} \boldsymbol{g}_0 \in\mathbb{C}^{N_a\times 1}.
\end{equation}

Based on the experienced channel and prior knowledge, it is assumed that each user can find and activate the port with the maximum channel gain. Let ${g}_k$ be the channel with the maximum gain from the $k$-th user to the base station, and the channel matrix between the $k$-th user and all receiving antennas at the base station is denoted by $\boldsymbol{g}_h$, which can be expressed as:
\begin{equation}
	\boldsymbol{g}_h = {g}_k\cdot[\boldsymbol h(1),\boldsymbol h(2),\dots,\boldsymbol h(M)]\in\mathbb{C}^{1\times M}.
\end{equation}

\subsection{Unsourced ISAC Signal Models}
The overall transmission is uniformly divided into $J$ independent chunks, where a two-phase transmission structure is adopted during each chunk. Specifically, the duration of each chunk consists of CS phase and FEC phase with $L_p$ and $L_c$ channel uses respectively, i.e., $L/J=L_p+L_c$. At CS phase, the CUs transmit the whole selected pilot signals and the SUs transmit the first segment pilot signals, for channel state information estimation including activity detection (AD), channel estimation (CE) and AOA estimation. At FEC phase, CU transmits the modulated signals and the SUs transmit the second segment pilot signals. Let $\boldsymbol{a}_i\in\mathbb{C}^{L_p\times 1}$ denote the first segment pilot signal of the $i$-th active SU and $\boldsymbol{b}_i\in\mathbb{C}^{L_c\times 1}$ denote the second segment pilot signal. $\boldsymbol{c}_j\in\mathbb{C}^{L_p\times 1}$ denote the selected pilot signal of the $j$-th active CU and $\boldsymbol{x}_j\in\mathbb{C}^{L_c\times 1}$ denote the $j$-th constellation frame of CU. The received signals at the CS phase $\boldsymbol{Y}_p\in \mathbb{C}^{L_p \times M}$ and the FEC phase $\boldsymbol{Y}_c\in \mathbb{C}^{L_c \times M}$ are expressed as:
\begin{equation}
	\label{eq:5}
	\begin{aligned}
		\boldsymbol{Y}_p &= \sum_{i=1}^{\bar{K}_s}{\boldsymbol{a}_i}\boldsymbol{G}_s(i)+\sum_{j=1}^{\bar{K}_c}{\boldsymbol{c}_j}\boldsymbol{G}_c(j)+\boldsymbol{N}_p,\\
		\boldsymbol{Y}_c &= \sum_{i=1}^{\bar{K}_s}{\boldsymbol{b}_i}\boldsymbol{G}_s(i)+\sum_{j=1}^{\bar{K}_c}{\boldsymbol{x}_j}\boldsymbol{G}_c(j)+\boldsymbol{N}_c,
	\end{aligned}
\end{equation}
where $\boldsymbol{N}_p$ and $\boldsymbol{N}_c$ are AWGN following distribution of $\mathcal{CN}(\boldsymbol{0},\sigma^2\boldsymbol{I})$, $\bar{K}_s=\mathbb{E}(K_s/J), \bar{K}_c=\mathbb{E}(K_c/J)$ are the average number of total active SUs and CUs.

\section{Encoder and Decoder Design}\label{sec.3}
\begin{figure*}[htbp]
	\centerline{\includegraphics[width=0.8\textwidth]{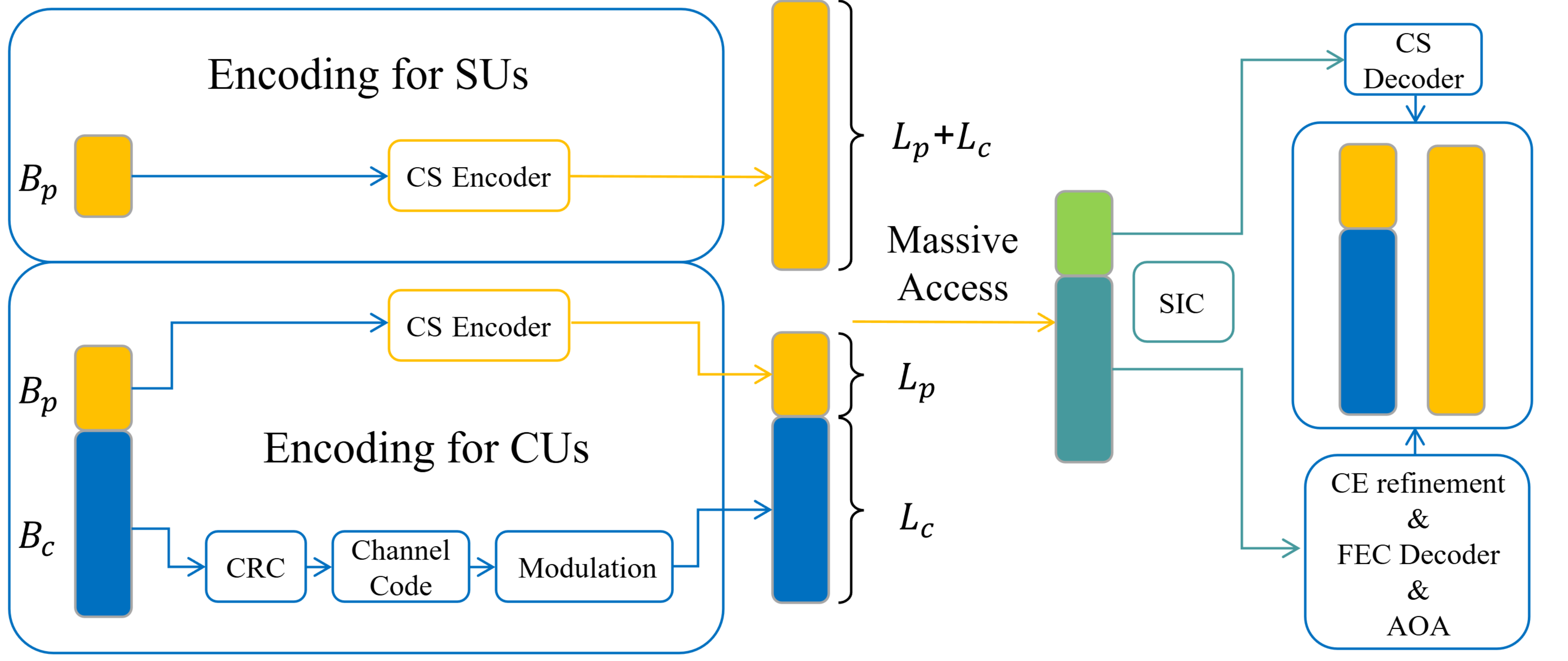}}
	\caption{Illustration of encoding and decoding procedures, where CE refers to channel estimation and SIC refers to successive
		interference cancellation.}
	\label{Encoding_Decoding}
	\vspace{-5mm}
\end{figure*}

The message of each UE includes chunk selection bits $B_\mathrm{chunk}$, pilot selection bits $B_p$ and information bits $B_c$. For CUs, $B=B_\mathrm{chunk}+B_p+B_c$; for SUs, $B_s=B_\mathrm{chunk}+B_p$. Let $\boldsymbol{c}_k\in \{0,1\}^{B\times 1}$ and $\boldsymbol{s}_k\in \{0,1\}^{B_s\times 1}$ denote the binary messages from CUs and SUs. UEs encode the binary messages based on their categories and the base station decodes the received signal to restore the $B$ bits of CUs and $B_s$ bits for SUs. We assume $K_c=K_s=K_a/2$ and the encoded message can occupy the whole chunk with $L/J$ length. The encoding and decoding process \cite{Slotted} is shown in Fig.~\ref{Encoding_Decoding}.
\subsection{Encoder Design}
\subsubsection{Encoding for SUs}
SUs first perform chunk selection using $B_\mathrm{chunk}$, and there are $J=2^{B_\mathrm{chunk}}$ chunks in total. After chunk selection, the user selects a pilot from a codebook generated by the discrete Fourier transform (DFT) method. Specifically, a large matrix is first generated via DFT. We select $2^{B_s}$ columns and $L/J$ rows from it to form the codebook $\boldsymbol{A}_s=[\boldsymbol{a}_{s,1},\boldsymbol{a}_{s,2},\boldsymbol{a}_{s,3},\dots,\boldsymbol{a}_{s,2^{B_s}}]\in\mathbb{C}^{L/J\times2^{B_s}}$ for SUs. Similarly, pilot codewords are selected by the integer $i_k,k\in[1:\bar{K}_s]$ generated from the $B_p$-bit binary messages, i.e., $\boldsymbol{a}_k=\boldsymbol{a}_{s,i_k}$. Let ($1-\alpha$) be the ratio of the sensing user’s power to the total transmit power. Pilot codeword abides by the power constraint of $\|\boldsymbol{a}_{s,i_k}\|_2^2 = P_s=(1-\alpha) P$. The projection from binary bits into pilot codeword is often known as CS encoding. Eventually, signal $\boldsymbol{a}_{s,i_k},k\in[1:\bar{K}_s]$ is transmitted over the first $L_p$ channel uses at the $j_k$-th chunk.
\subsubsection{Encoding for CUs}
The chunk selection and codebook selection method for the CUs are similar to those for the SUs. The difference is that the signal transmitted by the CU contains both the pilot sequence and the information sequence, where the length of the pilot sequence is $L_p$. Accordingly, the codebook for the CUs is constructed from different columns of the same DFT matrix as the SU codebook. For example, if the SUs codebook selects columns 1 to $2^{B_s}$, the CUs codebook selects columns $2^{B_s}+1$ to $2 \times 2^{B_s}$, which avoids pilot collision to the greatest extent.

For the $B_c$ information sequence of the CU, we first append CRC check bits, followed by channel coding, e.g., Polar code or low-density parity-check (LDPC) code, i.e., FEC encoding. The FEC-encoded vector is modulated with $Q$ order, e.g., for binary phase shift keying (BPSK), the modulation order is $Q = 2$, for quadrature phase shift keying (QPSK), $Q = 4$. The length of the modulated signal depends on the encoding scheme of Polar codes. If the code length is $E$, the signal length is $L_c=\frac{E}{\log_2 Q}$. Thus, the length of pilot for CU is $L_p=L/J-L_c$. Let $\beta$ be the ratio of the CU pilot power to the total CU transmit power. The pilot codeword follows the power constraint of $\|\boldsymbol{a}_{c,i_k}\|_2^2 = \beta P_c$ and the signal vector of constellation symbols follows the power constraint of $\|\boldsymbol{x}_k\|_2^2=(1-\beta)P_c$.
\subsection{Decoder Design}
The proposed decoder consists of four major components
working in tandem, including CS decoding, AOA estimation
for SUs, CE refinement and
channel code decoding. Moreover, SIC is adopted to enhance the overall performance.
\subsubsection{CS Decoding}
To detect the active users and estimate their corresponding channels from the received pilot signal, we employ the SOMP \cite{SOMP}, which is a greedy iterative algorithm to solve sparse
recovery problems. Its core lies in calculating the correlation (inner product) between the $\boldsymbol{A}_c$ and the signal in the CS phase to detect the indices of active CUs, and calculating the correlation between the $\boldsymbol{A}_s$ and the signal in the whole chunk to detect the indices of active SUs in each iteration.

After identifying the active users, the channel equalization technique based on minimum mean square error (MMSE) is employed in the algorithm. Let $\widetilde{\boldsymbol{A}}$ denote the set of pilots of active users in the codebook $\boldsymbol{A}_s$ or $\boldsymbol{A}_c$, and $\boldsymbol{Y}$ is the signal received. The estimated channel $\boldsymbol{H}$ is given by:
\begin{equation}
	\label{eq:CE}
	\widetilde{\boldsymbol{H}}=(\widetilde{\boldsymbol{A}}^{\mathrm{H}}\widetilde{\boldsymbol{A}}+\sigma^2 \boldsymbol{I})^{-1}\widetilde{\boldsymbol{A}}^{\mathrm{H}}\boldsymbol{Y},
\end{equation}
where $\widetilde{\boldsymbol{H}}=[\widetilde{\boldsymbol{h}}_1^{\mathrm{T}};\widetilde{\boldsymbol{h}}_2^{\mathrm{T}};\ldots;\widetilde{\boldsymbol{h}}_{\bar{K}_a}^{\mathrm{T}}]\in \mathbb{C}^{\bar{K}_a \times M}$ is the MMSE-based CE solution and $\widetilde{\boldsymbol{A}}$ contains all the codewords detected as active.
\subsubsection{AOA Estimation}
For AOA estimation, we employ the channel estimated in \eqref{eq:CE} to perform angle estimation, where the ESPRIT algorithm \cite{ESPRIT} is adopted. The principle of the ESPRIT algorithm is as follows. Consider a uniform linear array with equal spacing $d$. Array 1 selects $M_a$ antennas continuously, while Array 2 is the result of Array 1 shifted by one position. Thus, a phase delay $\Delta \theta$ exists between the two arrays for an incident signal $\boldsymbol{X}$:
\begin{equation}
	\boldsymbol{X}_2 = {\boldsymbol{X}_1}e^{j \Delta \theta}.
\end{equation}
Let the angle of incidence be $\boldsymbol{\widetilde\theta}_i$ (it is the angle relative to the array axis)
\begin{equation}
\cos \boldsymbol{\widetilde\theta }_i=-\frac{\Delta\theta\lambda}{2\pi d}.
\end{equation}
\subsubsection{CE Refinement}
To alleviate the error between the channel estimation based on OMP \cite{OMP} and the physical channel, we propose a CE refinement method, which leverages the inherent angular sparsity of the unsourced ISAC. Specifically, the channel estimation obtained from \eqref{eq:CE} is treated as the observation matrix $\tilde{\boldsymbol{H}}$, and we will refine the channel according to the actual array manifold $\boldsymbol{H}$ of the antenna array, which is regarded as dictionary.

The channel of FAS-ISAC consists of the array manifold $\boldsymbol{H}$ and the coefficient $\boldsymbol{g}$ following the complex Gaussian distribution, which can be regarded as a sparse signal reconstruction problem:
\begin{equation}
	\widetilde{\boldsymbol{H}} \approx\boldsymbol{H} \boldsymbol{g}.
\end{equation}
The sparse coefficient $\widetilde{\boldsymbol{g}}$ is solved by the OMP algorithm and channel refinement result $\boldsymbol{H}_r$ can be obtained.
\begin{equation}
	\boldsymbol{H}_r=\boldsymbol{H} \widetilde{\boldsymbol{g}}.
\end{equation}
\subsubsection{Successive Interference Cancellation (SIC)}
This paper introduces an alternating SU-CU Successive Interference Cancellation technique, designed to progressively eliminate interference from decoded users and subsequently re-evaluate the remaining users. Specifically, the system first employs initial channel estimation to equalize and perform Polar decoding on the communication users' data. Users with successful CRC verification are then filtered to form the set of perfectly decoded users. For these CRC-passed users, their reliable known pilots and information bits $\boldsymbol{X}$ are leveraged to conduct MMSE estimation on the original received signal, yielding a more accurate channel matrix denoted as $\boldsymbol{H}_\mathrm{SIC}$
\begin{equation}
	\boldsymbol{H}_\mathrm{SIC}=(\boldsymbol{X}^{\mathrm{H}} \boldsymbol{X}+\boldsymbol{I})^{-1}\boldsymbol{X}^{\mathrm{H}}\boldsymbol{Y}.
\end{equation}
Finally, these interferences are largely subtracted from the original received signal, providing a purer signal space for the decoding of SUs and the next round of CUs. Empirical testing reveals that under low-SNR scenarios, CUs require at least two rounds for complete detection. Therefore, the SIC ordering is configured as ``CUs-SUs-CUs".
\begin{table}[!t]
	\centering
	\caption{System Parameters}
	\begin{tabular}{ccc}
		\toprule
		\textbf{Variable} & \textbf{Symbol} & \textbf{Value}\\
		\midrule
		Number of CUs & $|\mathcal{L}_c|$ & ${K_a}/{2}$ \\
		Number of SUs & $|\mathcal{L}_s|$ & ${K_a}/{2}$ \\
		Chunk selection Bits & $B_{\text{chunk}}$ & 3 \\
		Pilot selection Bits & $B_p$ & 15 \\
		Ratios between users & $\alpha$ & 0.7 \\
		Ratios between phases & $\beta$ & 0.2 \\
		CRC length & / & 24 \\
		FEC phase length & $L_c$ & 512 \\
		CS phase length & $L_p$ & 113 \\
		\bottomrule
	\end{tabular}
	\label{tab:my_label}
\end{table}
\section{Numerical Results}\label{sec.4}

In this section, the performance of the proposed unsourced ISAC is
illustrated in terms of theoretical and empirical results.

\begin{figure}[t!]

	\centerline{\includegraphics[width=0.9\columnwidth]{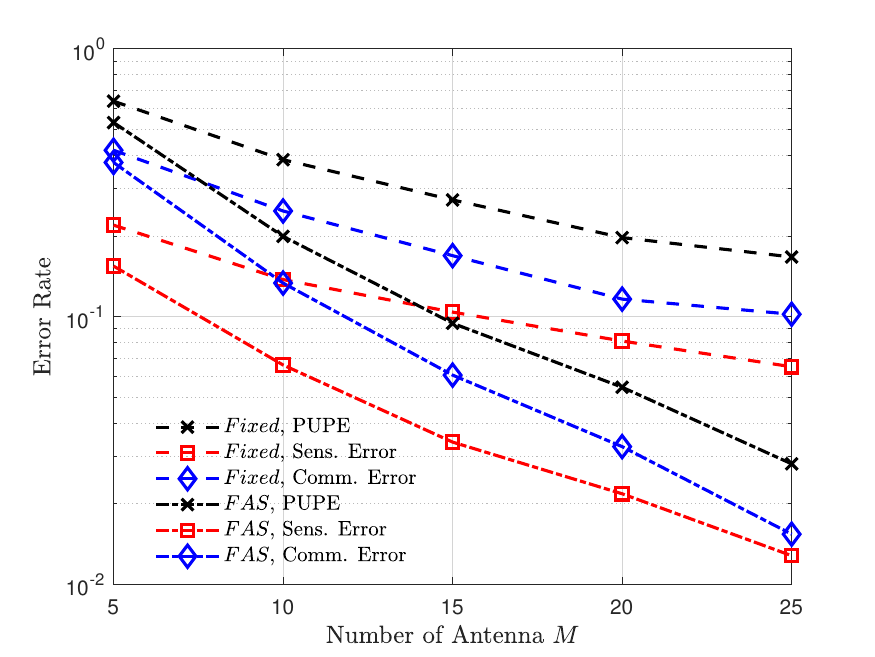}}
	\caption{Illustration of error rate versus different number of
		receiving antennas $M$ and antenna method with $K_a=800$ and $E/N_0$ = 10 dB.}
	\label{PUPE800}
	\centerline{\includegraphics[width=0.9\columnwidth]{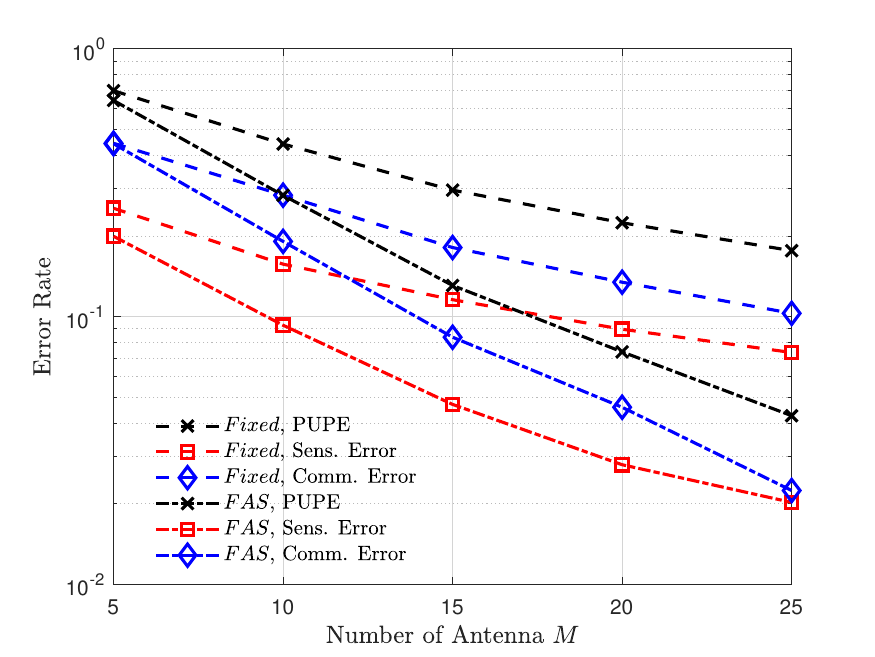}}
	\caption{Illustration of error rate versus different number of
		receiving antennas $M$ and antenna method with $K_a=1000$ and $E/N_0$ = 10 dB.}
	\label{PUPE1000}
	\centerline{\includegraphics[width=0.9\columnwidth]{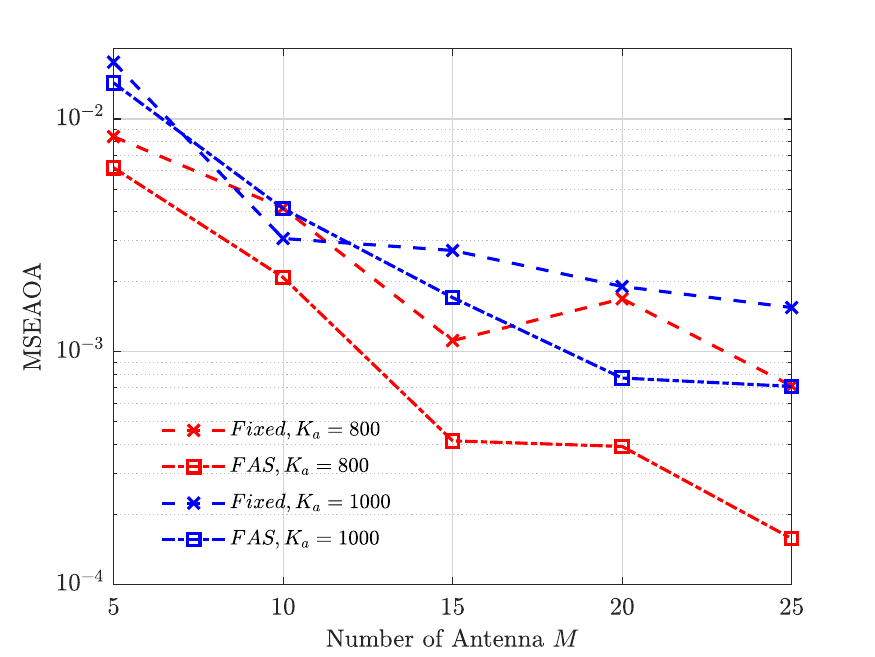}}
	\caption{Illustration of MSEAOA versus different number of
			receiving antennas $M$, antenna method and $K_a \in \{800,1000\}$ with $E/N_0$ = 10 dB.}
	\label{AOA}

\end{figure}
Overall, the parameter setups including channel uses $L = 5000$,
$B=100$ are fixed to align with the information theory
analyses. Other parameters are shown in Table~\ref{tab:my_label}. To reduce pilot collisions as much as possible, which would severely degrade the AOA estimation, the codebook size is set to a large value \cite{Pilot}. In practical engineering implementations, this may introduce substantial overhead.
\begin{figure}[t!]
	\vspace{-0.45cm}
	\centerline{\includegraphics[width=0.9\columnwidth]{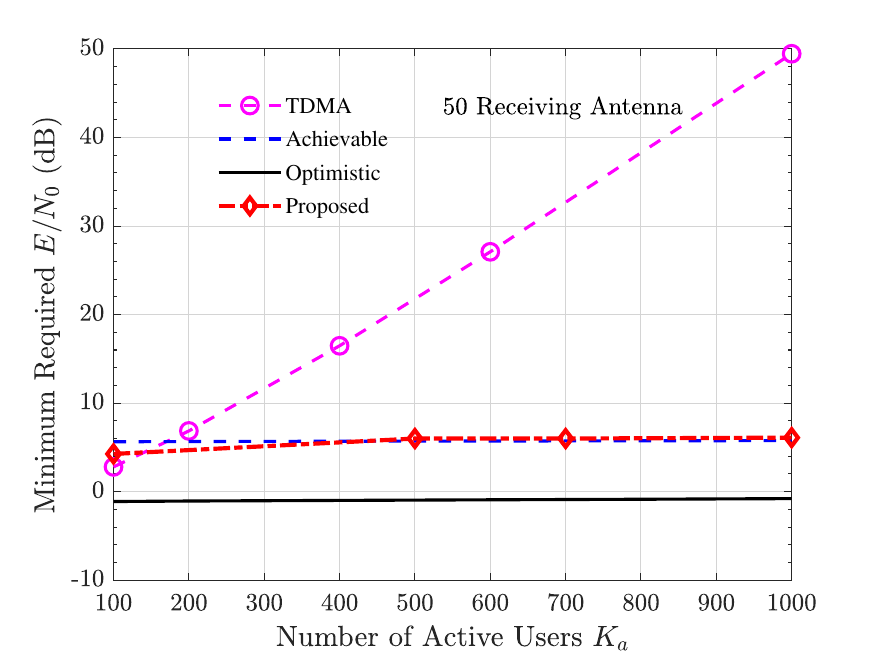}}
	\caption{Illustration of minimum required energy-per-user under small number
		of receiving antennas M = 50.}
	\label{EN050}
	\centerline{\includegraphics[width=0.9\columnwidth]{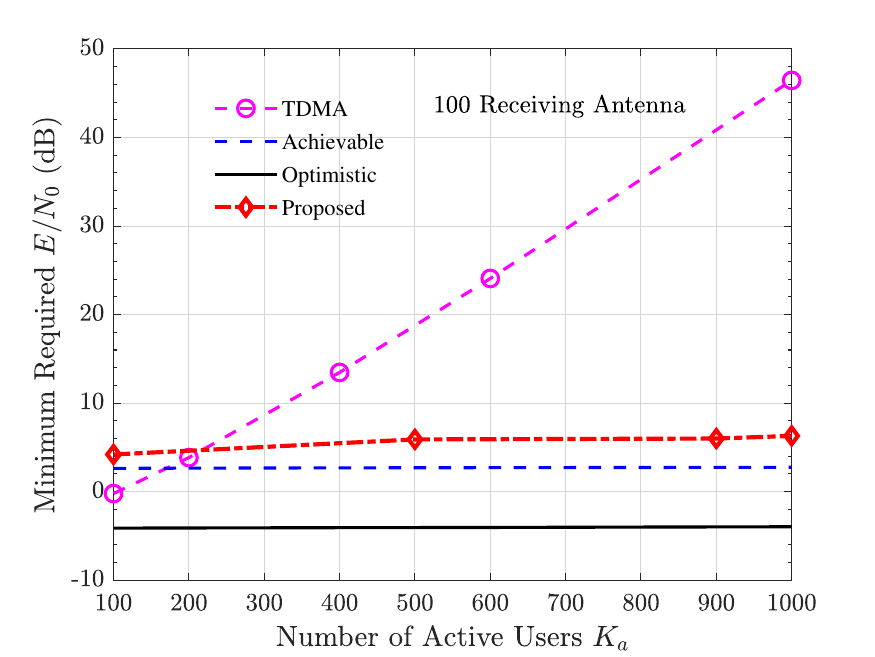}}
	\caption{Illustration of minimum required energy-per-user under small number
		of receiving antennas M = 100.}
	\label{EN0100}

\end{figure}

We illustrate different error rates including PUPE and its divisions of communication error and sensing error versus different number of antennas in Fig.~\ref{PUPE800} and Fig.~\ref{PUPE1000}, corresponding to $K_a=800$ and $K_a=1000$, respectively. Fig.~\ref{AOA} further reports the MSEAOA performance under these two user-load configurations. The ``\textit{Fixed~}'' in legend implies the scheme with a single fixed antenna deployed at the transmitter. Generally, both error rate and MSEAOA tend to decrease with increased number of receiving antennas. Meanwhile, error rate and MSEAOA generally tend to increase with increased number of users. It can be observed that in comparison with a single fixed antenna, the fluid antenna system can significantly reduce the error rate and enhance the accuracy of angle estimation.

Next, we evaluate the capacity performance under various configurations. The minimum required energy-per-user $E/N_0$ serves as the measure of capacity performance. The simulation includes a fluid antenna system with a length of $\lambda$ and benchmarks, which include TDMA, achievable and optimistic bounds in \cite{benchmark1}. Fig.~\ref{EN050} shows capacity performance with 50 received antennas. It should be noted that the achievable and optimistic bounds are derived for a single fixed antenna at the transmitter \cite{UNISAC3}. A single fluid antenna can find position with larger channel gain because of its higher degrees of freedom, to achieve this bound. When the number of received antennas increases to 100, the proposed scheme still approaches the achievable bound in Fig.~\ref{EN0100}. At 1000 active users, it is shown that the proposed scheme provides a 40 dB capacity gain over TDMA. These simulation results show that the proposed scheme is a novel approach with substantial performance improvements.

\section{Conclusion}\label{sec.5}
In this work, we propose a practical unsourced ISAC scheme design featuring slotted transmission. Compared to the recently established theoretical bounds for unsourced ISAC, the proposed scheme achieves system capacity close to the achievable bounds. In comparison with conventional multiple access schemes, the proposed scheme offers more than 40 dB of capacity gain with 1000 active users. Additionally, this paper proposes a novel scheme for user identification. The ``CUs-SUs-CUs" SIC pattern significantly enhances the detection success probability of active users. With the continuous maturity of fluid antenna hardware technology, FAS exhibits exceptionally promising prospects.

\end{document}